\documentclass[twocolumn]{aastex62}

\usepackage{xspace}

\newcommand{\secref}[1]{Section \ref{#1}}
\newcommand{\figref}[2]{Figure \ref{#1}{#2}}
\newcommand{\unit}[1]{\,\ensuremath{\mathrm{#1}}}
\newcommand{\Alfven}{Alfv\'en\xspace}

\received{ }
\revised{ }
\accepted{ }
\submitjournal{ApJL}

\shorttitle{Quasi-periodic pulsation}
\shortauthors{Yuan et al.}

\begin{document}

\title{A Compact Source for Quasi-Periodic Pulsation in an M-class Solar Flare}

\correspondingauthor{Ding Yuan}
\email{yuanding@hit.edu.cn}

\author[0000-0002-9514-6402]{Ding Yuan}
\affil{Institute of Space Science and Applied Technology,
	Harbin Institute of Technology, Shenzhen,
	Guangdong 518055, China}
\affil{Key Laboratory of Solar Activity, National Astronomical Observatories, Chinese Academy of Sciences, Beijing 100012, China}

\author[0000-0003-4709-7818]{Song Feng}
\affil{Yunnan Astronomical Observatory, Chinese Academy of Sciences, PO Box 110, Kunming 650011, China}
\affil{Yunnan Key Laboratory of Computer Technology Application, Faculty of Information Engineering and Automation, Kunming University of Science and Technology, Kunming 650500, China}
\author{Dong Li}
\affil{Key Laboratory for Dark Matter and Space Science, Purple Mountain Observatory, Chinese Academy of Sciences, Nanjing 210034, China}

\author{ZhongJun Ning}
\affil{Key Laboratory for Dark Matter and Space Science, Purple Mountain Observatory, Chinese Academy of Sciences, Nanjing 210034,  China}

\author{Baolin Tan}
\affil{Key Laboratory of Solar Activity, National Astronomical Observatories, Chinese Academy of Sciences, Beijing 100012, China}



\begin{abstract}
Quasi-periodic pulsations (QPP) are usually found in the light curves of solar and stellar flares, they carry the features of time characteristics and plasma emission of the flaring core, and could be used to diagnose the coronas of the Sun and remote stars. In this study, we combined the Atmospheric Imaging Assembly (AIA) on board the Solar Dynamics Observatory and the Nobeyama Radioheliograph (NoRH) to observe an M7.7 class flare occurred at active region 11520 on 19 July 2012. A QPP was detected both in the AIA $131\unit{\AA{}}$ bandpass and the NoRH $17\unit{GHz}$ channel, it had a period of about four minutes. In the spatial distribution of Fourier power, we found that this QPP originated from a compact source and that it overlapped with the X-ray source above the loop top. The plasma emission intensities in the AIA $131\unit{\AA{}}$ bandpass were highly correlated within this region. The source region is further segmented into stripes that oscillated with distinctive phases.  Evidence in this event suggests that this QPP was likely to be generated by intermittent energy injection into the reconnection region.
\end{abstract}

\keywords{Sun: atmosphere --- Sun: corona  --- magnetohydrodynamics (MHD) --- Sun: flare --- Sun: oscillations}

\section{Introduction} 
\label{sec:intro}

Quasi-periodic pulsation (QPP) is a rhythmic modulation to electromagnetic radiation of plasma in a wide range of frequencies during a solar or stellar flare. It is normally observed in integrated light curves in most electromagnetic bandpasses of solar and stellar plasma emissions, ranging from radio band, visible light, extreme ultraviolet (EUV) to X-rays \citep[see reviews of][]{Nakariakov2009,VanDoorsselaere2016}. Statistics with GOES soft X-ray emission suggests that 80\% of the solar flares exhibit QPP during the impulsive phase \citep{Simoes2015}. Since QPP carries the time characteristics of flare emission, it could be used to diagnose the key parameters of flaring site  \citep[e.g.,][]{Brosius2015,Pugh2019}. 

Two theories are proposed to explain the origin of QPP in a flare: repetitive magnetic reconnection \citep{McLaughlin2009,McLaughlin2012,Thurgood2017,McLaughlin2018,Dominique2018} and modulation to the flaring site by magnetohydrodynamic (MHD) waves \citep{Nakariakov2006,Reznikova2011,Tian2016}. Repetitive magnetic reconnection cannot be observed directly,  since coronal magnetic field is extremely difficult to measure with current instrumentation. Moreover, the dynamics of a current sheet, which is a direct byproduct of magnetic reconnection, does not normally reveal itself in the EUV emissions of coronal plasma. Albeit difficult, several studies attempted to estimate the magnetic field or current sheet with lucky viewing angles in imaging or spectroscopic measurement \citep{Su2013, Longcope2018, Warren2018}. To investigate the physics of QPP, one has to seek for signature of repetitive waves and flows during magnetic reconnection \citep{McLaughlin2012}. 

If one suggests the modulation by MHD waves, a plausible step is to correlate the timestamps in integrated light curves and periodic waves. A series of studies was done on modulation by standing and reflective slow mode waves \citep{Wang2003b,Wang2003a,Wang2011,Kumar2013,Kumar2015,Yuan2015,Fang2015,Mandal2016}. Most previous studies focus on a good match of periodicity and phase difference. However, after nine years' launch of the Solar Dynamics Observatories \citep[SDO,][]{Pesnell2012}, although its onboard instrument - the Atmospheric Imaging Assembly \citep[AIA,][]{Lemen2012} has sufficient spatial and temporal resolutions in nine UV and EUV channels, no observation provides convincing evidence on the causality between MHD wave and QPP. Therefore, the attempts to diagnose the reconnection site of solar and stellar flares with QPPs are weakly supported \citep[e.g.,][]{Kim2012}. 

A correct model is extremely nontrivial for stellar flare QPP, since the spatial resolution is not achievable for a remote star. Once the nature of QPP is justified, it would become a novel tool to diagnose stellar atmospheres. In this study, we found that QPP originates from a very compact source above the loop top, and that no connectivity was found between QPP and other oscillatory signals at the lower atmosphere. This letter is structured as follows: \secref{sec:obs} presents data preparation and methods; \secref{sec:results} describes the main results and is followed by a discussion in \secref{sec:con}.

\section{Data reduction and analysis}  
\label{sec:obs}
\begin{figure*}[ht]
\includegraphics[width=\textwidth]{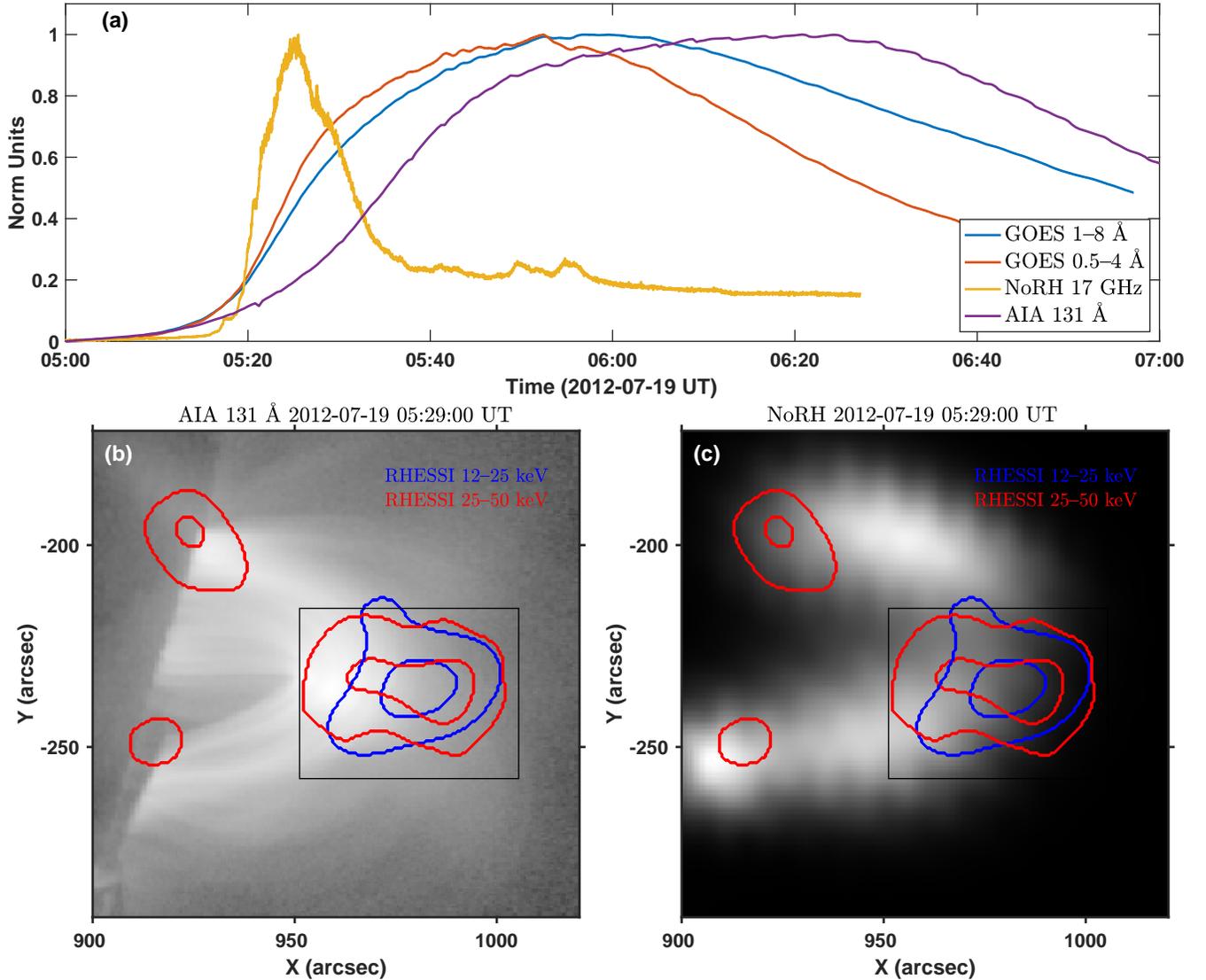}
\caption{An arcade of loops at AR 11520 recorded by the AIA $131\unit{\AA{}}$ channel (b) and NoRH $17\unit{GHz}$ band (c) at 06:10 UT on 19 July 2012. The RHESSI X-ray emission intensity were overlaid with contour levels at 50\% and 90\% of the peak intensity. Two energy bands were computed at 12-25 keV (blue) and 25-50 keV (red), respectively. (a) Light curves of the GOES SXR flux at 1-8 \AA{}, 0.5-4 \AA{}, the NoRH radio flux at $17\unit{GHz}$, and the AIA $131\unit{\AA{}}$ bandpasses. They are normalized to the peak value so that the maximum value is one. The AIA $131\unit{\AA{}}$ signal was averaged over the rectangle at the loop top as marked in panel (b).   \label{fig:fov}}
\end{figure*}

\begin{figure*}[ht]
\includegraphics[width=\textwidth]{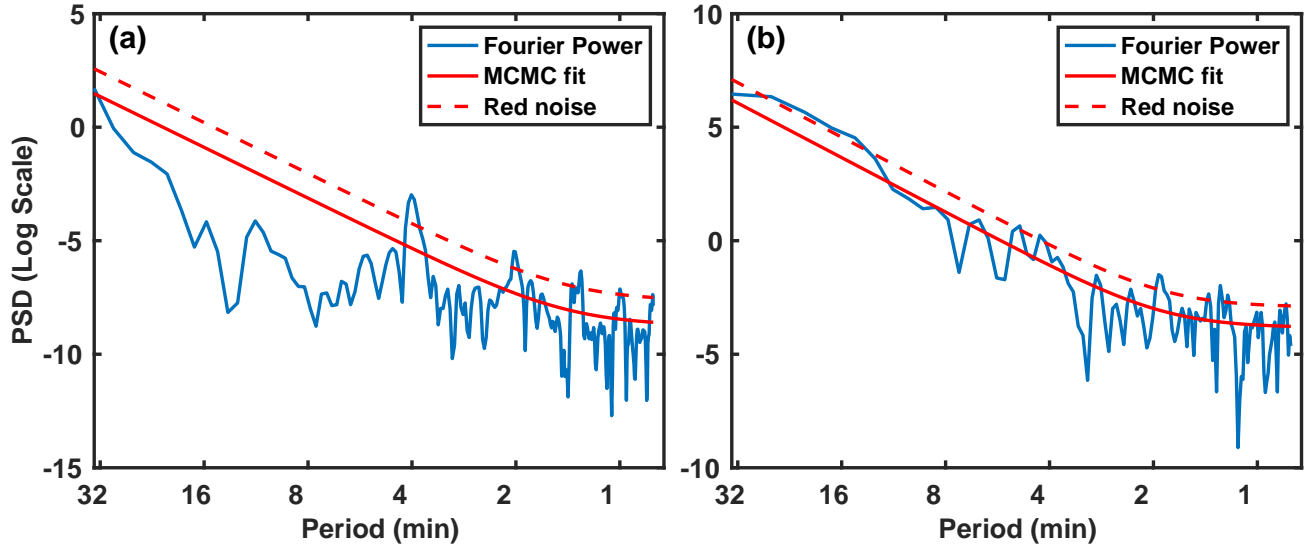}
\caption{Fourier power spectra of the light curves of the AIA $131\unit{\AA{}}$ bandpasses (a) and the NoRH radio flux at $17\unit{GHz}$ (b) as shown in \figref{fig:fov}(a). The Markov chain Monte Carlo (MCMC) fit and red noise  at 95\% confidence level (or 5\% significance level) are over-plotted as guideline for assessing significant oscillatory signals. \label{fig:powerlaw}}
\end{figure*}

\begin{figure*}[ht]
\includegraphics[width=\textwidth]{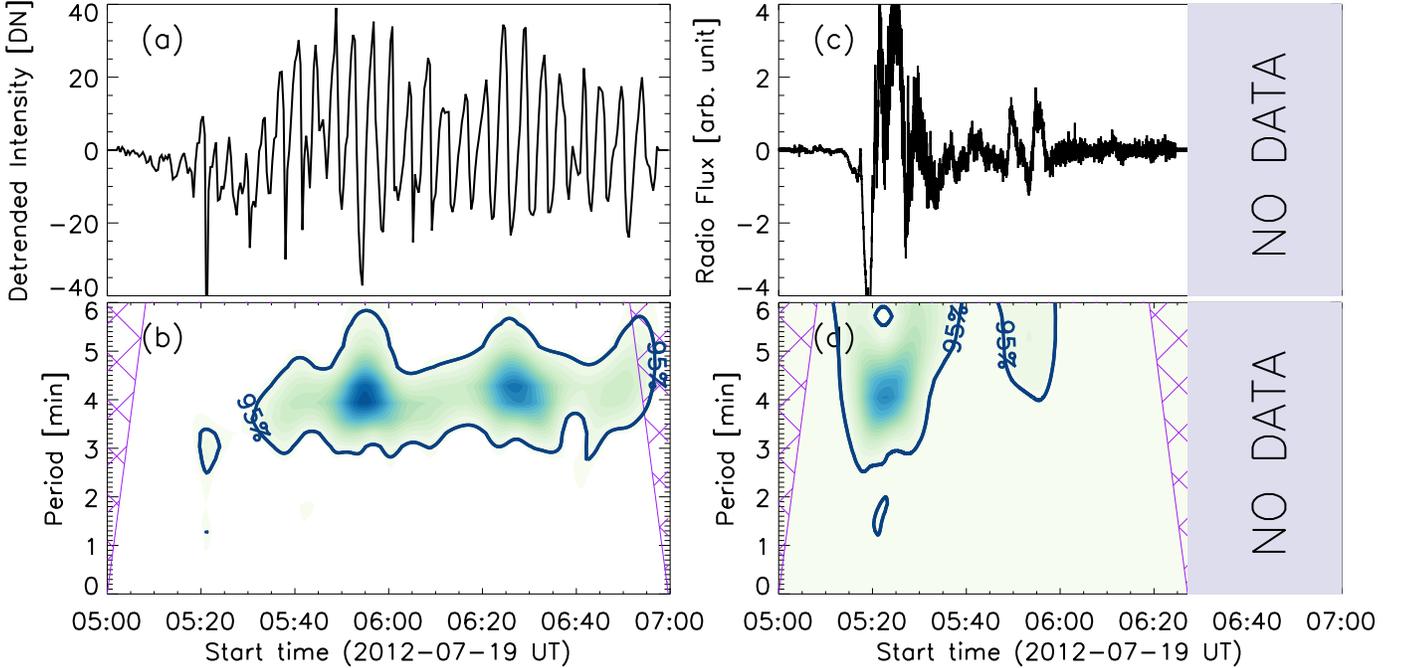}
\caption{(a) Light curve of the average $131\unit{\AA{}}$ signal after removing the five minute moving average. (b) Wavelet power spectrum. (c) and (d) are the same analysis to the NoRH $17\unit{GHz}$ radio flux. The contours in (b) and (d) are the 95\% confidence level (or 5\% signifcance level). The cones-of-influence are cross-hatched, they represents the meaningless power affected by zero-paddings.  
	\label{fig:wavelet}}
\end{figure*}

\begin{figure*}[ht]
\includegraphics[width=\textwidth]{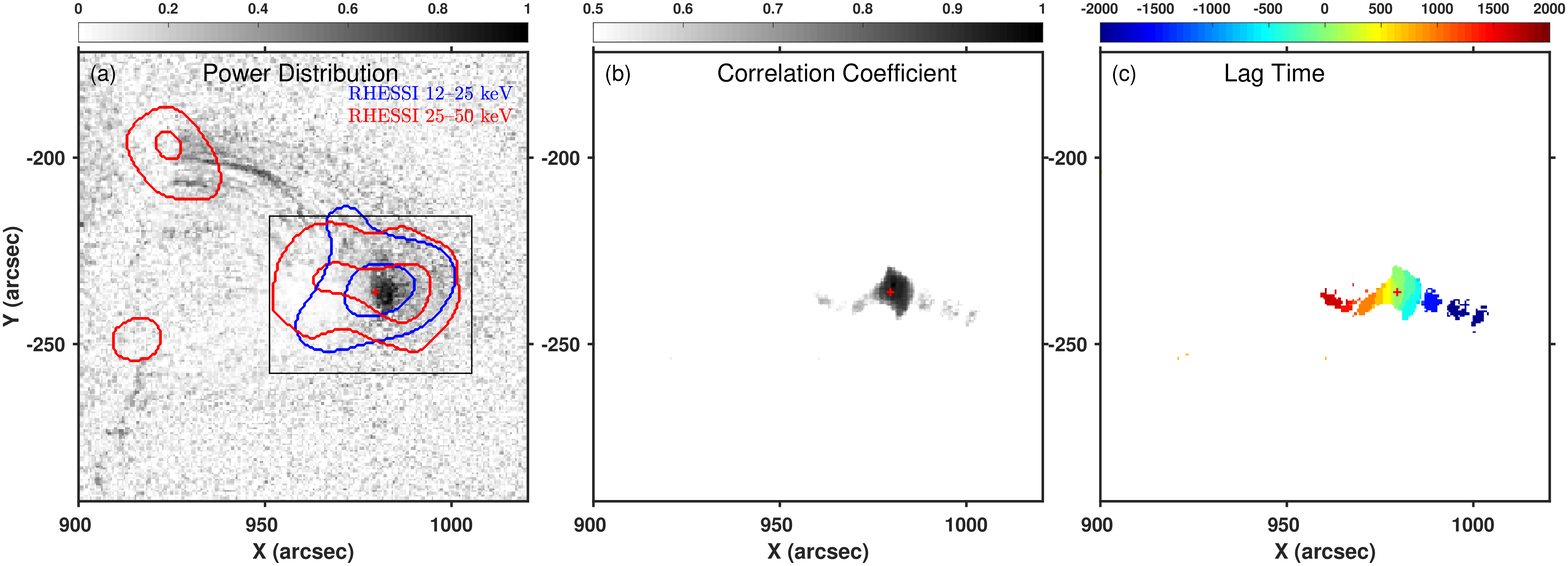}
\caption{(a)-(c) Maps of Fourier power averaged between 200-300s, correlation coefficient and lag time. The red plus symbol represents the reference pixel. Panel (a) only draws the pixel with four minute oscillation above the $3\sigma$ white noise level. The contours in (a) is the same as plotted in \figref{fig:fov}.  \label{fig:powermap}}
\end{figure*}

On 19 July 2012, A GOES class M7.7 flare was detected on NOAA active region (AR) 11520. This AR has rotated to the west limb of the sun, an arcade of coronal loops was well oriented for observation in both EUV and radio emission bandpasses (\figref{fig:fov}{b} and \ref{fig:fov}{c}). \figref{fig:fov}{a} presents the Soft X-ray fluxes recorded by the Geostationary Operational Environmental Satellite (GOES), the average EUV emission intensity captured by the AIA $131\unit{\AA{}}$ channel, and the average radio flux at $17\unit{GHz}$ received by the Nobeyama Radioheliograph \citep[NoRH,][]{Nakajima1994}. The Soft X-ray flux started to raise at about 4:17 UT, reached a peak value at about 6:00 UT and decays off gradually. The NoRH $17\unit{GHz}$ radio flux rapidly raised to peak value and decayed off  from about 5:15 UT to about 5:40 UT. \figref{fig:fov}{a} reveals QPPs in the Soft X-ray fluxes and the average EUV emission close to the peak time of this flare. No apparent QPP was visible in the NoRH $17\unit{GHz}$ signal, however we will show later that significant periodic signal was buried in this light curve.

The magnetic structure of this active region was well exposed for investigating solar flare models and its dynamics at outer corona, several article were published to report various aspect of this flare and the associated dynamics.  \citet{LiuW2013} measures the plasmoid ejections and seek for evidence of particle acceleration during magnetic reconnection;  \citet{Sun2014} studies the thermal structure of flaring loops; \citet{LiuR2013} investigate the currents behind an erupting flux rope; \citet{Patsourakos2013} reported evidence of a fast coronal mass ejection driven by the prior information of a magnetic flux rope. \citet{Wu2016} used Nobeyama Radioheliograph data to study the non-thermal electron emissions, they detected an intermittent time scale of about two minute for magnetic reconnection; More specifically, \citet{Huang2016} reported quasi-periodic acceleration of electrons that may coincide in the bi-direction outflows from the reconnection region.

SDO/AIA data have excellent spatial and temporal resolutions to reveal the source and time characteristics of this QPP. The AIA data were calibrated with the standard calibration routine provided by the SolarSoft package\footnote{\url{http://www.mssl.ucl.ac.uk/surf/sswdoc/solarsoft/}}. The digital offset for the cameras, CCD readout noise and dark current were removed from the data, then each image was corrected with a flat-field and was normalized with its exposure time.  The saturated images were disregarded, so the cadence is reduced to $24\unit{s}$ in the EUV data. NoRH recorded the $17\unit{GHz}$ radio flux till 6:27 UT, the cadence was fixed to $1\unit{s}$. 

We used the Reuven Ramaty High Energy Solar Spectroscopic Imager \citep[RHESSI,][]{Lin2002} detectors 3 and 5-9, and the CLEAN algorithm to compute the X-ray images. The measurement started on 5:29:30 UT and integrated for $30\unit{s}$. The contours of X-ray emission intensity were overlaid in Figures \ref{fig:fov}{b}, \ref{fig:fov}c, and \ref{fig:powermap}{a}. 

\figref{fig:fov}{a} plots the EUV emission intensity variation in the AIA $131\unit{\AA{}}$ channel, this signal was averaged within the area enclosed by the black rectangle (\figref{fig:fov}{b}). We estimated the global trend by taking the five minute moving average, and removed it from the light curve to highlight the periodic pulsations. 

Some researchers are concerned about the effect of de-trending, so they developed a robust spectral method to detect oscillatory signals in the original light curves \citep[e.g.,][]{Gruber2011,Inglis2015,Pugh2017}.  These algorithms could reveal significant oscillatory signals within a Fourier spectrum, which normally follows a power-law distribution. This practice could appropriately account for the red noise in the data \citep{Pugh2017}. We calculated the Fourier spectra with original light curves of both the $131\unit{\AA}$ and $17\unit{GHz}$ fluxes, and assessed the significance level with red noise distribution and Markov chain Monte Carlo (MCMC) fit (see \figref{fig:powerlaw}). This results confirmed that the four minute oscillation analysed throughout this study is above the 95\%-confidence level in both the AIA $131\unit{\AA}$ and NoRH $17\unit{GHz}$ signal. We also note a significant spectral peak with two-minute periodicity, but its amplitude is very small, so it is hardly detectable in wavelet spectrum (see \figref{fig:wavelet}). We hereafter opt to wavelet analysis with detrended times series, so that the temporal behaviour and the connectivity between different observables could be examined.

The detrended time series was plotted in \figref{fig:wavelet}{a} and was analysed with wavelet transform \citep{Torrence1998}, the wavelet spectrum was plotted in \figref{fig:wavelet}{b}. Similar analysis was done to the NoRH $17\unit{GHz}$ data,  the time series and wavelet spectrum were plotted in \figref{fig:wavelet}{c} and \ref{fig:wavelet}d, respectively. 

To study the spatial distribution of QPP, we perform Fourier transform to the detrended emission intensity of every pixel in the AIA $131\unit{\AA{}}$ and average the Fourier power among the spectral components between $200\unit{s}$ and $300\unit{s}$ (\figref{fig:powermap}{a}). This range includes significant periodic signal revealed in \figref{fig:wavelet}{b}. Before this step, we removed the five minute moving average in each time series as done in the wavelet analysis. Within each pixel, we estimate the white noise level for the Fourier power spectrum according to \citet{Torrence1998}. \figref{fig:powermap}{a} only presents the pixels  with four minute oscillation power above the $3\sigma$ noise level. In the NoRH $17\unit{GHz}$ images, the loop top emission was very weak (\figref{fig:fov}). The plasma density and temperature was significantly enhanced above the loop top \citep{Huang2016, Wu2016}, therefore, the opacity for the $17\unit{GHz}$ emission varies between footpoints and loop top. So similar analysis to the NoRH dataset could not effectively reveal the spatiotemporal knowledge about the source above the loop top. Our event is different with the case studied in \citet{Inglis2008}, in which the coronal loops had a very different orientation  \citep[see Figure 6 in ][]{Inglis2008}.   

We note that the region with QPP was very compact and only localized above the loop top. To reveal the inter-correlation within the source region, we calculated the cross-correlation coefficient (XC) of the detrended emission intensity of each pixel and that of a reference pixel at the loop top as labelled in \figref{fig:powermap}. We measured the lag time by finding the argument maximum of XC for each pixel, the maximal XC and lag time within the region of interest are illustrated in \figref{fig:powermap}{b} and \figref{fig:powermap}{c}.

\section{Results} 
\label{sec:results}
During this M7.7 flare, we observed several QPPs with distinct periods in the Soft X-ray, radio and EUV emission light curves, \citet{Huang2016} reports a thorough study on the quasi-periodic processes. In this study, we focus on the spatio-temporal characteristics of QPP in the EUV imaging data.  This sort of study was attemptd with NoRH radio observation, e.g., \citep{Inglis2008,Fleishman2008}, however, the coarse spatial resolution of radio observation is  not sufficient to reveal the fine structure of QPP. With a thorough understanding of QPP in solar flares, we could apply the knowledge to stellar flares with QPP and derive more details of a remote star.

\figref{fig:fov}{a} shows that in the averaged $131\unit{\AA{}}$ light curve quasi-periodic signals was clearly visible from about 5:30 to 7:00 UT. \figref{fig:wavelet}{a} and \ref{fig:wavelet}{b} represents respectively the de-trended signal and its wavelet power spectrum. A significant period is detected at about four minutes, the periodicity last persistently for about 90 minutes. We also detected an oscillatory signal with the same periodicity in the NoRH $17\unit{GHz}$ radio signal (\figref{fig:wavelet}{c} and \ref{fig:wavelet}{d}), however, this QPP only lasted from 5:15 UT to 5:40 UT. It occurred about 15 minutes ahead of the EUV signal;  two oscillatory signals had an overlap of about 10 minutes duration (or 2.5 oscillatory cycles). 

\figref{fig:powermap}{a} presents the Fourier power distribution of the four minutes periodicity. We found that this four minutes oscillation was only significant above the loop top, and its spatial extent was compact and had overlap with hard X-ray emission region. Moreover, these oscillations were highly correlated, the XC coefficient was greater than 0.5 in the compact region above loop top, see \figref{fig:powermap}{b}. \figref{fig:powermap}{c} reveals that this oscillation could be segmented into several sub-regions. Within each sub-region the lag time was uniform. However, between two neighbouring sub-regions, the lag time had a difference of about $200\unit{s}$ - $300\unit{s}$, this is very close to a oscillation cycle. 

\section{Discussions}
\label{sec:con}

In this study, an arcade of loops at AR 11520 was well oriented for observation. An M7.7 flare was detected on 19 July 2012 at this AR. During this flare, QPP was detected both in the light curves of AIA $131\unit{\AA{}}$ bandpass and the NoRH $17\unit{GHz}$ radio flux. The emission intensities oscillated with a period of about four minutes. 

The QPP had a very compact source and was only localized above the loop top. Its location had overlap with the hard X-ray source above the loop top. The QPP source region exhibited high correlation and was segmented into several stripes according to the distribution of lag time. 

The oscillatory signal in the NoRH $17\unit{GHz}$ started about 15 minute ahead of the AIA $131\unit{\AA{}}$ signal. The order of start time of two detected oscillatory signals implies that this QPP might be triggered by an energetic process, which was first observed in high energy radiation and then detectable  in the EUV emissions. The QPP in the AIA $131\unit{\AA{}}$ channel did not exhibit significant damping during a 90-minutes interval. Similar persistent QPPs were detected at the impulsive phase and decay phase of an X8.2 solar flare  \citep{Hayes2019}. This feature is contradictory to those QPPs with strong damping \citep[e.g.,][]{Anfinogentov2013}. So this kind of QPP should be driven periodically or by a sequence of impulsive energy injections. 

If we consider the periodicity at MHD timescale and its persistence, such a signal resembles the leakage of sunspot oscillations \citep{Yuan2011, Yuan2016,Li2018}. However, we did not observe similar oscillatory signal propagating along the loops in the EUV emission intensity (see \figref{fig:powermap}{a}), so we rule out the possibility of leakage of sunspot oscillation in forms of slow mode wave. If such leakage is the cause, the propagation of energy should not disturb the density or temperature of the flaring loop, then it could be non-thermal particles or \Alfven waves. \citet{Chen2006} demonstrated that solar $p$-mode could modulate the reconnection site and generate a oscillatory signal. However we did not found sufficient evidence to support this theory in this event.

This QPP is also likely to be caused by self-consistent periodic modulation to the energy releasing site at the loop top by colliding flows or non-thermal plasma. Such a scenario is reproduced by MHD simulations \citep{Fang2016,Ruan2019}.	\citet{Ruan2019} simulated a turbulent flare loop by triggering counter-stream flows at loop top and reproduced a compact QPP source with  highly correlations. 

Another possible origin is the intermittent bombardment of quasi-periodic plasmoid generated by magnetic reconnections. This process was numerically predicted by \citep{Takasao2016,Zhao2019}. The event studies in this letter is more inclined to support this origin. \citet{Wu2016} detected a  two minute periodicity in the pre-impulsive phase of the same flare, this periodicity is about half of the period measured in this study. In our study, we also detect significant  two minute oscillation, but its amplitude was very small compared to the four-minute oscillation (see \figref{fig:powerlaw}{} and \figref{fig:wavelet}). The two minute oscillations in the impulsive and decay phase might be connected to the signal in the pre-impulsive phase as presented in \citet{Wu2016} , but we don't have conclusive evidence at this stage. Such a rhythmic process is accompanied by fast contracting loop and upward ejective plasmoids \citep{LiuW2013}. This scenario appears to be consistent with the lag time distribution (\figref{fig:powermap}{c}). 

In this study, we report both the spatial extent and causality of a QPP. Although the physical mechanism of QPP is not conclusive, we obtained a wealth of new features about QPP. It would be a good start for future investigations. A combination of imaging and spectroscopic study should be able to get insight into the origin of this kind of QPP.

\acknowledgments
We would like to thank the anonymous referee for his or her helpful comments. D.Y. is supported by the National Natural Science Foundation of China (NSFC, 11803005, 11911530690), Shenzhen Technology Project (JCYJ20180306172239618), and the Open Research Program (KLSA201814) of Key Laboratory of Solar Activity of National Astronomical Observatory of China. S.F. is supported by the Joint Fund of NSFC (U1931107) and the Key Applied Basic Research program of Yunnan Province (2018FA035). D.L. and Z.J.N are supported by the NSFC (11973092，11603077, 11573072) and the Youth Fund of Jiangsu (BK20171108).
Wavelet software was provided by C. Torrence and G. Compo, and is available at URL: \url{http://atoc.colorado.edu/research/wavelets/}. The authors would like to acknowledge the ISSI-BJ's international workshop on ``Oscillatory Processes in Solar and Stellar Coronae''. The data were provided by the SDO/AIA, RHESSI, and Nobeyama teams.

\bibliography{yuan2019qpp}


\end{document}